\documentclass[i12pt,preprint]{aastex}
\begin{document}
\title{\bf Detecting the Baryons in Matter Power Spectra}

\author{Christopher J. Miller, Robert C. Nichol}
\affil{Department of Physics, Carnegie Mellon University, 5000 Forbes Ave.,
Pittsburgh, PA 15232, USA (chrism@cmu.edu \& nichol@cmu.edu)}

\author{Xuelei Chen}
\affil{Institute for Theoretical Physics, University of California, 
Santa Barbara, CA 93106, USA (xuelei@itp.ucsb.edu)}

\begin{abstract} 
We examine power spectra from the Abell/ACO rich cluster survey
and the 2dF Galaxy Redshift Survey (2dFGRS) for observational evidence of features 
produced by the baryons. A non-negligible baryon
fraction produces relatively sharp oscillatory 
features at specific wavenumbers in the
matter power spectrum. However, the mere existence of baryons will also produce
a global suppression of the power spectrum. We look for both of these
features using the false discovery rate (FDR) statistic. 
We show that the window effects on the Abell/ACO power spectrum are
minimal, which has allowed for the discovery of discrete oscillatory 
features in the power spectrum.
On the other hand, there are no statistically significant
oscillatory features in the 2dFGRS power spectrum, 
which is expected from the survey's broad window function.
After accounting for window effects we apply a scale-independent
bias to the 2dFGRS power spectrum, $P_{Abell}(k) = b^2P_{2dF}(k)$ and $b = 3.2$.
We find that the overall shapes
of the Abell/ACO and the biased 2dFGRS power spectra are entirely consistent
over the range $0.02 \le k \le 0.15h$Mpc$^{-1}$.
We examine the range of $\Omega_{matter}$ and baryon fraction, for which
these surveys could detect significant suppression in power.
The reported baryon fractions for both the Abell/ACO and 2dFGRS surveys are
high enough to cause a detectable suppression in power (after accounting for errors,
windows and $k$-space sampling). 
Using the same technique, we also examine, given the best fit baryon
density obtained from BBN, whether it is possible to detect additional
suppression due to dark matter-baryon interaction.  
We find that the limit on dark matter cross section/mass derived from 
these surveys are the same as those 
ruled out in a recent study by Chen, Hannestad and Scherrer.
\end{abstract}


\section{Introduction}

During its first $\simeq100,000$ years, the
Universe was filled with a fully ionized plasma with a tight coupling between the photons
and baryons via Thomson scattering. A direct consequence of this coupling is
the acoustic oscillation of both the primordial temperature and
density fluctuations (within the horizon) caused
by the trade--off between gravitational collapse and photon pressure.  The
relics of these acoustic oscillations are predicted to be visible as alternating
peaks and valleys in the CMB temperature power spectrum. 
The relative amplitudes and locations of the features
provide powerful constraints on the cosmological parameters ({\it e.g.}
$\Omega_{total}, \Omega_{b}$, etc).
The BOOMERANG,
MAXIMA, and DASI experiments announced the first high confidence detection of these
acoustic oscillations in the temperature power spectrum of the Cosmic
Microwave Background (CMB) radiation (Miller et al. 1999; Melchiorri et al. 2000;
Balbi et al. 2000; Lee et al. 2001;
Netterfield et al. 2002; Halverson et al. 2002;
de Bernardis et al. 2002; Miller et al. 2002a).

Like the CMB experiments mentioned above, we have also
seen the recent emergence of large and/or deep
extragalactic datasets to directly measure the 3-dimensional luminous matter
power spectrum, $P(k)$.
Specifically, the PSCz galaxy survey (Saunders et al. 2000), the Abell/ACO
cluster survey (Miller et al. 2002b), and the
2dF Galaxy Redshift Survey (Colless et al. 2001) have all released
newly measured power spectra
(Hamilton, Tegmark, and Padmanabhan, 2000; Hamilton and Tegmark 2002;
Miller \& Batuski  2001; Miller, Nichol \& Batuski 2001a,b; Percival et al. 2001;
Tegmark, Hamilton \& Xu 2002).
These surveys are beginning to allow us to search for oscillatory
features in the matter power spectrum
(frozen into the matter distribution from acoustic oscillations 
on the largest-scales).
Using some of these large (in volume and/or number) surveys, Miller et al. (2001a)
found statistically significant features (dips in power)
at $k \sim 0.04$ and $0.1h$Mpc$^{-1}$ in the $P(k)$.
While much less obvious, features (at 
similar wavenumbers) were also qualitatively observed
in the recently determined 2dFGRS $P(k)$ (Percival et al. 2001) and the 2dF QSO
power spectrum (Hoyle et al. 2002). 
The existence of such oscillatory features in the matter power 
spectrum permits one only to
say whether they are consistent with baryonic features, since
features could also be due to other reasons (e.g. features in the primordial $P(k)$).

Besides the features resulting from acoustic oscillations,
the presence of a non-negligible baryon fraction in the Universe has another
observational consequence on the power spectrum:
the overall power below the sound horizon is suppressed over a large range in $k$
(see Eisenstein and Hu 1998--hereafter EH98).
The suppression occurs in both the cold dark matter and baryonic 
transfer functions. As discussed in EH98, the main effect of the baryons is the
suppression of the growth rates between the equality and drag epochs.
Therefore, the effect of baryons in the matter power spectrum can be detected via
this suppression over many $k$-modes, as opposed to looking for the sharp features
which result from the acoustic oscillations. 

However, there are some ``degeneracies'' in the $P(k)$ suppression.
In addition to the baryonic damping discussed above, 
dark matter (hereafter--DM) scattering off the baryons will also suppress the power
spectrum on scales smaller than 
the sound horizon (Chen, Hannestad, and Scherrer 2002), and 
hot dark matter will also have the effect of suppressing power on small scales. 
Eisenstein and Hu (1999)
note that the suppression of the transfer function due to  baryons will dominate
over neutrinos.  Others have begun to address the effects of hot dark matter
(see Wang, Tegmark, and Zaldariagga 2002 and Elgaroy et al. 2002),
and so we focus here on suppression effects related to the baryons: baryon suppression
and DM-baryon scattering.

In Section 2 we compare the 2dFGRS power spectrum to that of the Abell/ACO cluster
spectrum, taking into account effects from their 
respective window functions. By controlling
the False Discovery Rate (Miller et al. 2001a,c), we look for any statistically
significant oscillatory features in the 2dFGRS power spectrum. In Section 3, we then examine
whether either of these surveys could have detected the baryons through suppression.
In Section 4, we examine the effect of 
DM--baryonic scattering interactions on the $P(k)$. We then summarize and discuss
our findings in Sections 5 and 6.

\section{The 2dFGRS and Abell/ACO $P(k)$}

\subsection{The Surveys and their Spectra}
In this work, we have chosen to examine two of the
largest luiminous-mass samples available today: the Abell/ACO
cluster catalog and the 2dF Galaxy Redshift Survey
catalog. The cluster catalog covers 2$\pi$ steradian
(the entire sky excluding $|b| < 30^{\circ}$) and has over 600 clusters to $z = 0.14$.
This sample is presented in Miller et al. (2002b) and the power spectrum
can be found in Miller \& Batuski (2001).
The 2dFGRS covers a much smaller portion of the sky (a southern strip
$85 \times 15^{\circ}$ and a northern strip $75 \times 10^{\circ}$).
However, the 2dFGRS power spectrum probes to a deeper redshift
than the Abell/ACO ($z = 0.25$) and has $\sim 150,000$ galaxy redshifts.
The sample selection and definition for the 2dFGRS can be found in 
Colless et al. (2001) and the power spectrum in Percival et al. (2001).
We chose these samples for their size (in volume and in number) and
also because their $P(k)$ were calculated using the Feldman, Kaiser,
and Peacock (1994) prescription. Also, as we show below, after accounting
for the window effects on the 2dFGRS and relative bias, the shapes of
the two power spectra are indistinguishable.

\begin{figure}
\plotone{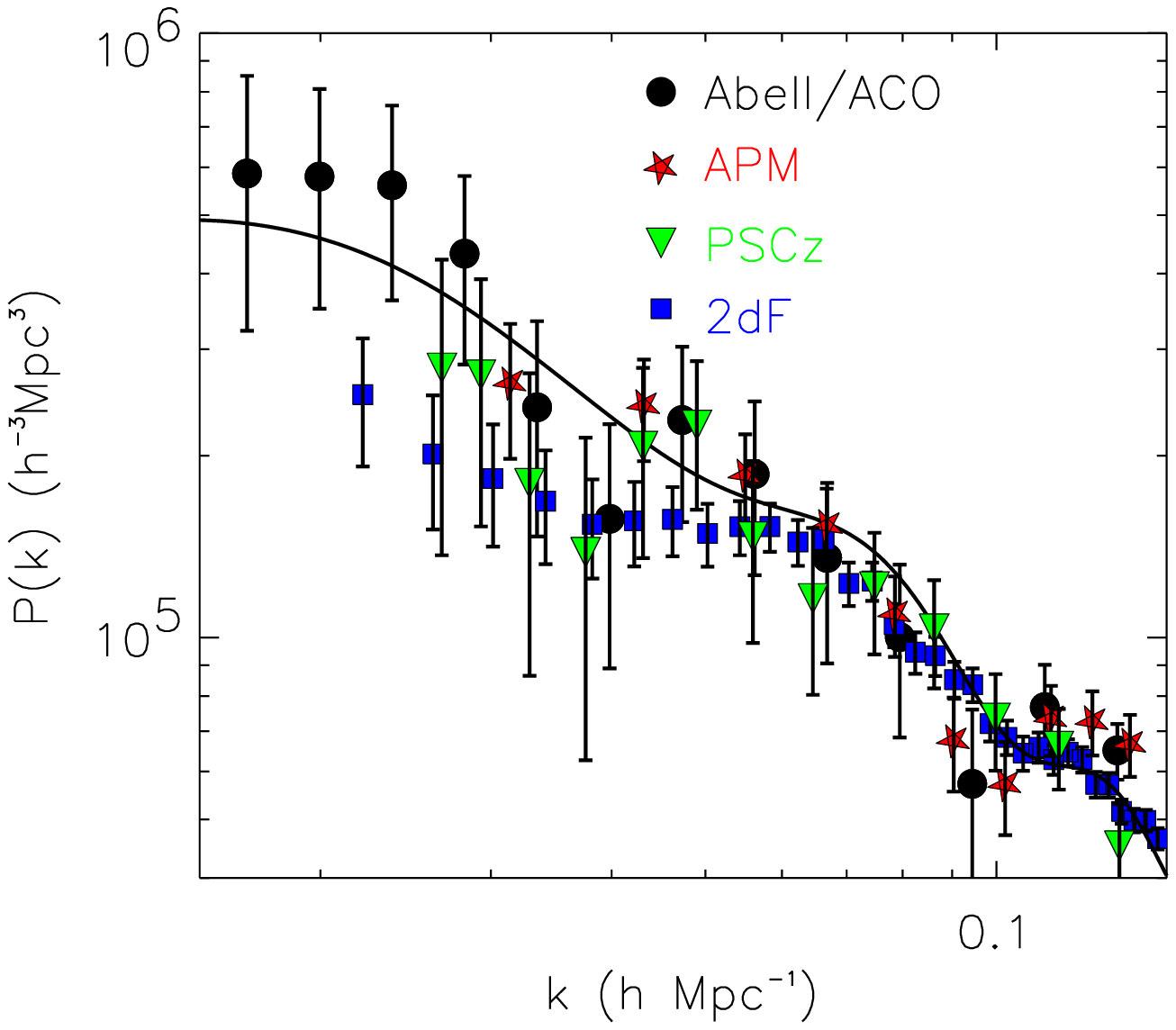}
\caption[]{The power spectra for four surveys of clusters and galaxies.
The triangles are the PSCz galaxies, the stars are the APM groups
and clusters, the circles are the Abell/ACO rich clusters, and the squares
are the 2dF galaxies. All power spectra have been normalized to the Abell/ACO.
This constant relative bias
corresponds to
$b = 1.5$, $b = 3.6$, and $b = 3.2$ for
Abell/ACO versus APM, Abell/ACO versus the PSCz, and Abell/ACO versus 2dFGRS respectively.
The features seen at $k \sim 0.04$ and $k \sim 0.1h$Mpc$^{-1}$ are significant
in the APM, Abell/ACO and PSCz, but not in the 2dFGRS.
}
\label{fig::ps_many}
\end{figure}

In Figure \ref{fig::ps_many}, we plot the $P(k)$ of 
the three surveys used in Miller et al. (2001a), along with
the 2dFGRS measurement of Percival et al. (2001). In this plot, all of
the $P(k)$ have been normalized to the Abell/ACO amplitude over the 
range $0.07 \le k \le 0.12h$Mpc$^{-1}$.
The features (dips) seen in the Abell/ACO, PSCz and APM measurements
are not nearly as evident in the 2dFGRS data.
In Figure  \ref{fig::ps_many}, one should notice that for $k > 0.05 h$Mpc$^{-1}$
the shapes of all four spectra are nearly identical. We discuss the difference
at larger scales ($k < 0.05h$Mpc$^{-1}$) below.

\subsection{Detecting Features and Deviations}

Throughout this work, we will be performing statistical
tests to differentiate between some given null hypothesis (e.g.
a spectrum without baryonic oscillations) and the data (real or modelled).
Instead of looking at the overall shapes of two power spectra (using a
$\chi^2$ test, for example), we
approach this problem from a multiple-hypothesis testing point-of-view.
We choose this type of statistical technique because we will want to detect 
either (a) discrete features in the spectrum for which a $\chi^2$ test provides
very little information (like their location),
or (b) the global
suppression of the $P(k)$, and not simply strong features caused by acoustic oscillations.
For instance, a multi-hypothesis testing approach allows
us to demand that all $P(k)$ measurements for some $k > k_{min}$
are rejected, and not just the bumps and dips.

Our choice of testing procedure will be via controlling the
False Discovery Rate or FDR
(see Miller et al. 2001a and
Benjamini \& Hochberg 1995).
FDR is a simple procedure which allows one to control the number of mistaken
rejections (or false discoveries) when performing multiple
hypothesis tests. One pre-specifies the acceptable fraction of
false discoveries as 
$\alpha = \langle \frac{\rm False ~ rejections}{\rm Total ~ rejections} \rangle$,
then the FDR procedure defines an appropriate significance
threshold so that an error rate  $\leq \alpha$ is obtained for 
the data in question.  
For example, if $\alpha = \frac{1}{3}$ 
and we found in our tests that six of the data points are rejected from the null
hypothesis 
(in this case, the null hypothesis is these points being drawn from 
a smooth, featureless power spectrum),
then on average, only two of these points are erroneous rejections 
of the null hypothesis. 
A key advantage of the FDR procedure is that
it works with highly correlated data, such as the Percival et al. 2dFGRS $P(k)$.
While we do not have the 2dFGRS covariance matrix, we do not need it to apply
the FDR procedure (at the cost of a slightly more conservative statistical test).

The FDR technique has since be applied in a variety of astrophysical
scenarios. 
For instance, Miller et al. (2001c) 
applied the FDR technique to source detection algorithms in astronomical images. 
Hopkins et al. (2002) then extended on this work by combining
source pixels into extended sources and studying the final error rates.
Recently, Tago et al. (2002)
used FDR to search for oscillatory features in the large-scale
structure two-point correlation function on scales $> 200 h^{-1}$Mpc.

Operationally, the FDR technique requires that we first compute the $p$--value for each
result obtained in the tests. \footnote{The $p$--value 
is the probability that a random sampling
would lead to a data value with an equal or higher deviation from the
the null hypothesis.} Herein we have used
a smooth CDM model based on the Percival et al. (2001)
best-fit cosmological parameters,
but with the baryon oscillations removed (EH98).
We rank, in increasing size, the $p$-values of
each data and plot these as points $(j/N, p_j)$, where $j$ is the
rank of the point. We then 
draw a line of slope $\beta = \frac{\alpha}{Log\,N}$ and zero intercept,
where $N$ is the number of points and $Log\,N$ 
accommodates for correlated data.  The numerator, $\alpha$ is the user
specified false discovery rate. The
first crossing of this line with a $p$--value (moving from larger to smaller
$p$--values) defines the significance threshold $\sigma_{rej}$, below which
all points are rejected based on our null hypothesis. On average, only
$\alpha\times100\%$ of these rejected points will be mistakenly identified
as significant (or rejected by the test) as a result of
random fluctuations of the null hypothesis.

\subsection{Oscillatory Features and Window Effects in the 2dFGRS}

In this section, we apply the FDR tecnhique to 
determine whether the measured values of the 2dFGRS $P(k)$ are
consistent with a smooth, featureless,
underlying power spectrum.  Miller et al. (2001a), performed such an analysis on
the Abell/ACO, PSCz, and APM cluster data (shown in Figure \ref{fig::ps_many}). They
found features at $k \sim 0.04$ and $k \sim 0.1$.
In fact, we do not find any statistically significant
features in the 2dFGRS $P(k)$, even though qualitatively one sees
dips and valleys in the $P(k)$ from Percival et al. (2001) (see Figure 1).

Next, we examine what effect the windows of these
surveys have on their measured $P(k)$. We ask whether a feature
as strong as that seen in the Abell/ACO data (and also seen in the PSCz $P(k)$
from Hamilton and Tegmark 2002) could have been detected through the
window of the 2dFGRS.
As we shall show below, the effect of window ``smearing'' is
significant for the 2dFGRS.

\begin{figure}
\plotone{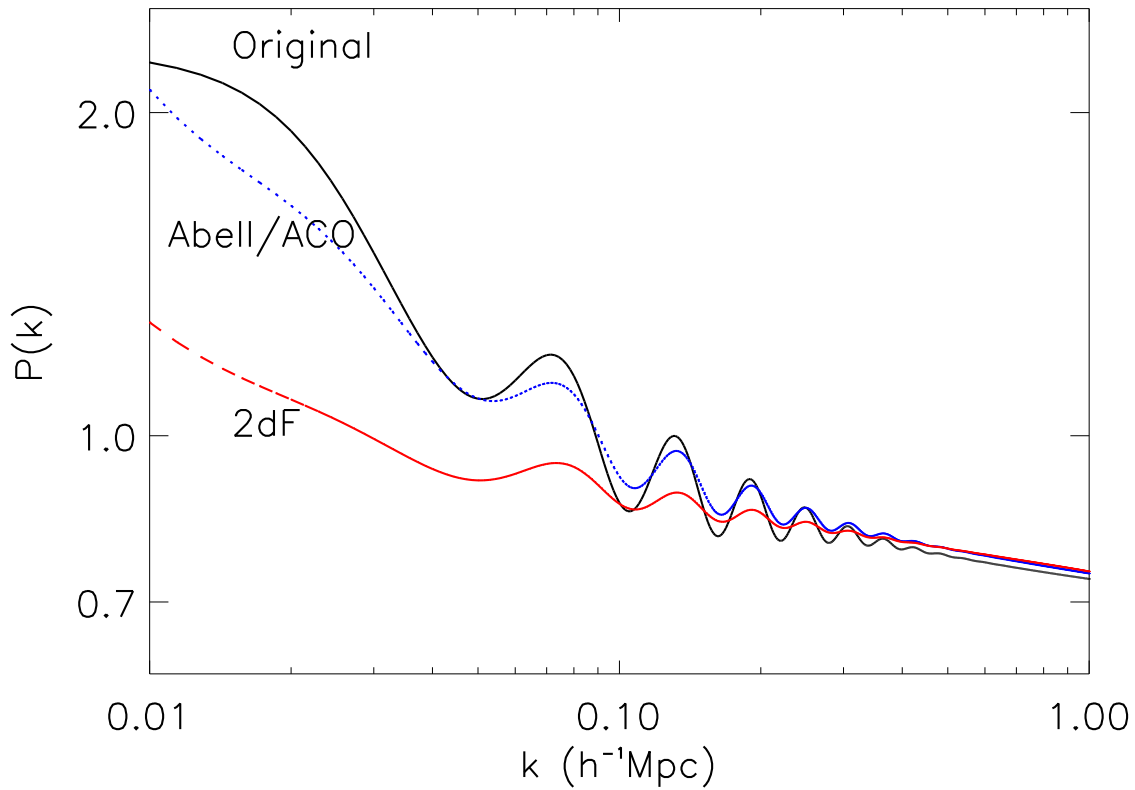}
\caption[]{The effects of the window functions on the 2dFGRS and Abell/ACO samples.
The power spectrum is for $\Omega_mh^2 = 0.14$, $\Omega_bh^2 = 0.029$, and $n_s = 1.08$
divided by a zero-baryon model. Note how the convolution of the power with the 
Abell/ACO window has little effect on the oscillatory features, while the 2dFGRS window
smears those features considerably.
}
\label{fig::window}
\end{figure}

In Figure \ref{fig::window}, we show the effects of the Abell/ACO window function
compared to the 2dFGRS window function. In this Figure, 
we use the cosmological model parameters
from Miller et al. (2001a,b), which have oscillatory features resulting from a
non-zero baryon fraction ($\Omega_mh^2 = 0.14$, $\Omega_bh^2 = 0.029$, and $n_s = 1.08$)
To be consistent with Percival et
al. (2001), we divide the baryonic power spectrum
by a zero-baryon model ($\Omega_m = 0.2$ and $\Omega_b = 0$) which has the effect
of visually enhancing any features.
We then convolve this model $P(k)$ with the Abell/ACO 
window (from Miller \& Batuski 2001) and also the 2dFGRS window
(given in Percival et al. 2001).
The difference between the height of the unconvolved peak and the Abell/ACO convolved
peak at $k = 0.07h$Mpc$^{-1}$ is $\sim 5\%$. The difference in the locations of the
maximum of the unconvolved and Abell/ACO convolved peaks is $< 5\%$. However, the 
2dFGRS convolution reduces the height of the peak by $\sim 50\%$ and shifts the peak
towards smaller $k$ by $\sim  20\%$. The effect from the 2dFGRS window gets smaller
as we move towards larger $k$.

\begin{figure}
\plotone{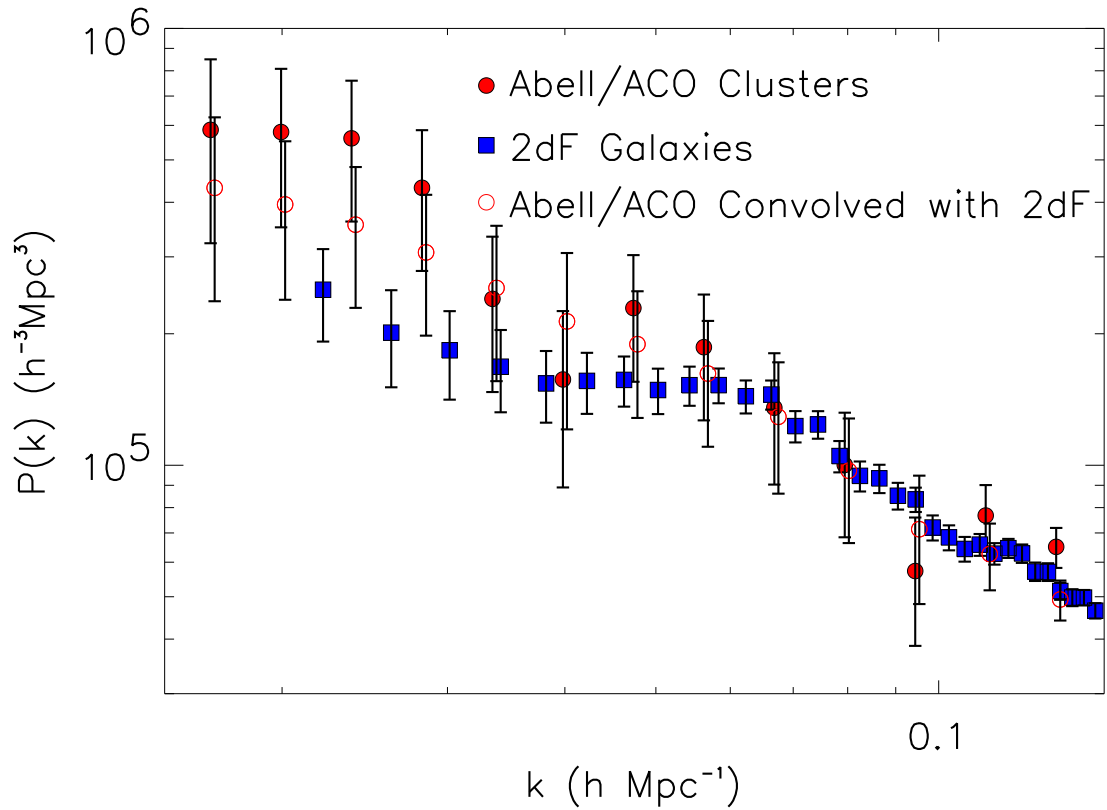}
\caption[]{The Abell/ACO cluster $P(k)$ after convolving with the 2dFGRS window. Note
that the features seen in the Abell/ACO $P(k)$ are washed out by the 2dFGRS window function.
}

\label{fig::abe_convolve}
\end{figure}

Figure \ref{fig::window} shows that while the
Abell/ACO window has minimal effect (i.e. $< 5\%$) on the amplitudes of the features,
the 2dFGRS window will smear the features significantly. We then can convolve
the Abell/ACO power spectrum of Miller \& Batuski (2001) with the window of the 2dFGRS
(under the approximation that the Abell/ACO spectrum is entirely unaffected by the
its window). In Figure \ref{fig::abe_convolve}, we see that the features in the
Abell/ACO $P(k)$ are entirely washed out. If we now perform the same statistical
search for features in the Abell/ACO $P(k)$, we do not find any features.

An advantage of surveys like the PSCz and Abell/ACO are their 
approximate full-sky coverage, resulting in a well
determined and highly peaked window function.  While 2dFGRS has 
a large number of points (galaxies), it also has
a very broad window (from its irregular shaped volume). Therefore, while the
2dFGRS $P(K)$ will have smaller error bars, the oddly shaped survey volume of the 2dFGRS
makes any statistical search for features difficult, if not impossible.

When comparing the convolved Abell/ACO and the 2dFGRS galaxy power spectrum
(shifted to match the amplitude of the clusters), we see that the overall shapes of the 
two are consistent (within their errors)
on scales $\sim 50 - 300h^{-1}$Mpc ($0.02 \le k \le 0.15h$Mpc$^{-1}$).
A Kolmogorov-Smirnoff test finds no significant difference between these two
power spectra.
Note that neither the 2dFGRS nor the Abell/ACO power spectra appear to turn-over at
large scales. As discussed by Miller \& Batuski (2001), this is especially 
important for the Abell/ACO survey since the window function has little
effect on the power spectrum at near gigaparsec scales ($k \sim 0.01$). A possible
cause of excess power on large-scales is from leakage of power on smaller
scales due to the broadness of the window function. However, Miller \& Batuski 
have shown that such leakage should be minimal for the Abell/ACO survey for
$k \sim 0.015$. The fact that we are not seeing a turnover in any of the
new large-surveys could imply an ever decreasing $\Omega_mh$ (so long
as the baryonic fraction of matter is not large and the dark matter is cold).

A remarkable aspect of Figure \ref{fig::abe_convolve} is the 
overall success of a simple linear
biasing model in re-normalizing the amplitudes of these power spectra over
nearly a decade of scale. A similar result was already found for the Abell/ACO, PSCz,
and APM surveys in Miller et al. (2001a). This 
scale--independent biasing model, over the range in $k$
discussed herein, has already been predicted in recent numerical simulations
(Narayanan, Berlind \& Weinberg 2000).
We have applied techniques to calculate the
amplitude shift ({\it e.g.} a $\chi^2$ minimization of the data with
identical $k$-values as well as using model fits to the data and
re--normalizing them to the Abell/ACO data), but in all cases we obtain
identical results: the relative bias (where $P_{Abell} = b^2P_{2dF}$)
between the Abell/ACO cluster sample
and the 2dF galaxy power spectra is $b= 3.2$ 
over the range $0.02 \le k \le 0.15h$Mpc$^{-1}$.

\subsection{Summary of the $P(k)$ Shapes}
In this section, we showed that the shapes of the 2dFGRS and
Abell/ACO are indistinguishable after accounting for
relative bias and window effects. We also showed that the
features seen in the 2dFGRS $P(k)$ are not significant.

In their likelihood analyses, Percival et al. (2001) find a non-negligible baryon 
fraction in two localized regions of parameter
space: ($\Omega_mh, \Omega_b/\Omega_m) \simeq (0.2,0.15)$ and $(0.6, 0.4)$.
These two regions are likely a result
of degeneracy in the models (see Percival et al. 2001). The
high baryon fit has very strong features, while the low baryon model is a better
overall fit to the 2dFGRS power spectrum, but still has features. 
However, since we find no observed features in the 2dFGRS
power spectrum, we examine the effects
of baryon suppression. Specifically, we ask at what baryon fraction will
the Abell/ACO and 2dFGRS surveys be able to differentiate between
a zero-baryon model and a baryon model (with the same $\Omega_m$).

\section{Suppression via the Baryons}

In this section, we present a method for detecting the baryons
based on comparing a non-baryonic model to the data. On the
largest scales (and all other parameters being equal), a baryonic
and non-baryonic power spectrum will have the same shape and
amplitude for a given dataset. One could then normalize
the model (with no baryons) to the data on these very large scales,
and search for the baryonic suppression on smaller scales (e.g $k > 0.03$).
However, the current measurements of power on the largest scales are not
precise enough to warrant such an analysis (the fractional errors on the measured
power spectra for the Abell/ACO and 2dFGRS surveys are $>30\%$ on the largest
scales).
So instead of the measured $P(k)$ for these surveys,
we use model power spectra (while still using the errors
on the measured $P(k)$) and vary the baryon fraction
to conjecture on the detectability of baryon suppression.

A possible source of concern in our analysis is non-linear
evolutionary effects which are known to erase the baryonic
oscillations. However,
these non-linear effects are not expected to be significant for
$k < 0.2h$Mpc$^{-1}$ (Mieksin, White, and Peacock 1999).
Also, Percival et al. (2001)  used
fully evolved N-body simulations (see their Figure 4)
to show that non-linear effects do not strongly alter
the recovered linear power spectrum for $k < 0.15h$Mpc$^{-1}$.
Therefore, throughout the rest of this work, we only use
power spectrum measurements for $k < 0.15h$Mpc$^{-1}$.

\begin{figure}
\plottwo{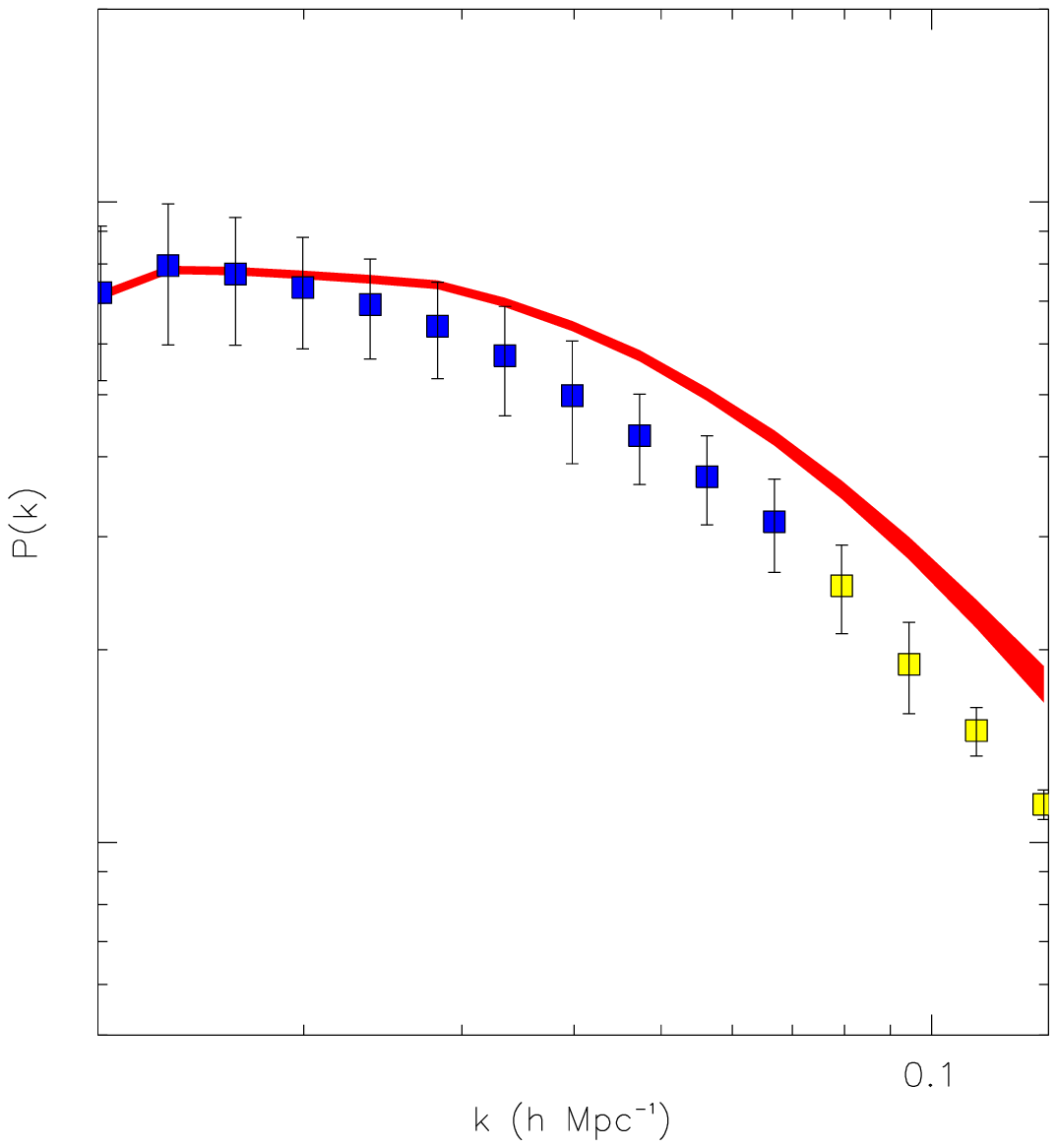}{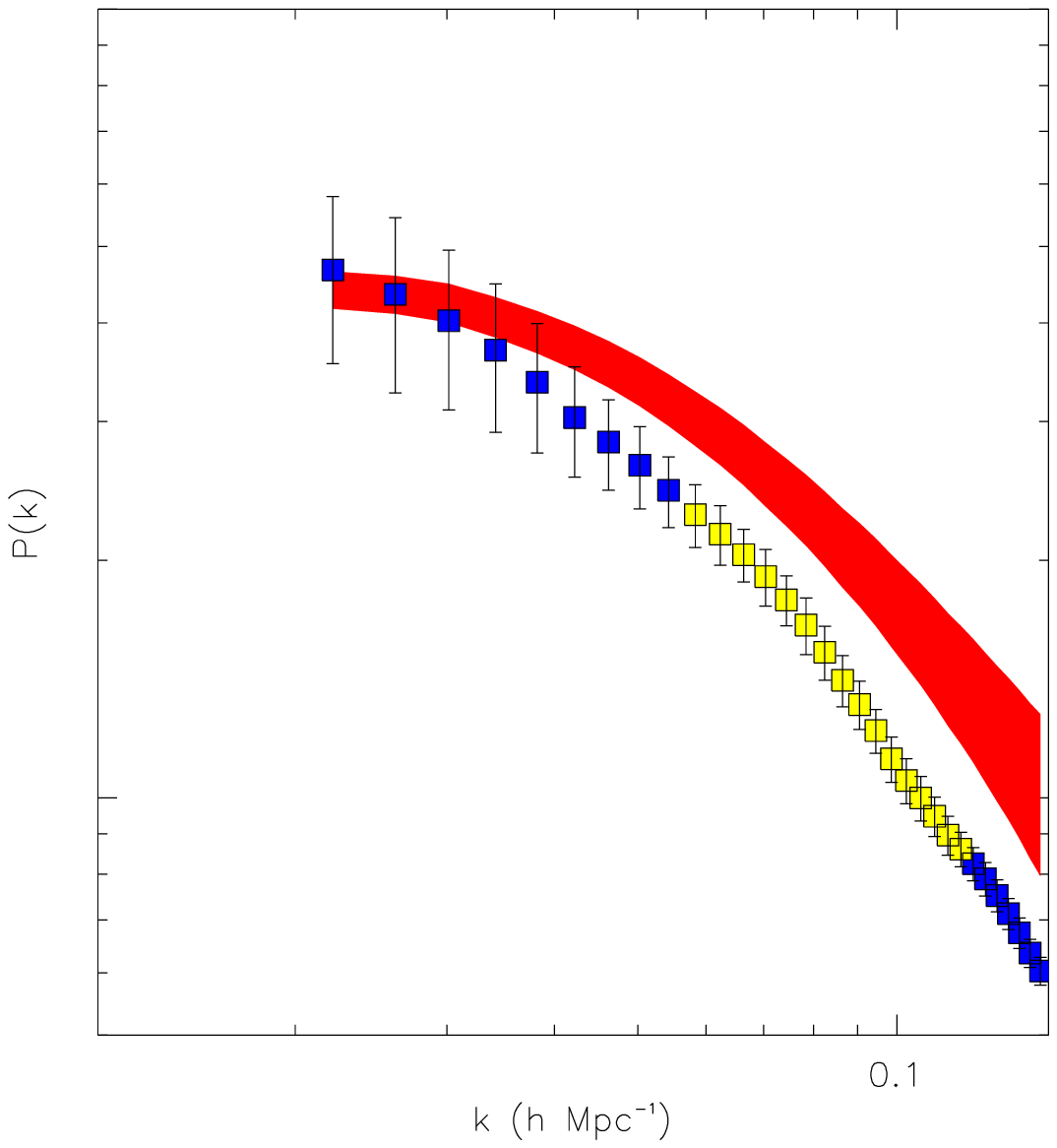}
\caption[]{An example of baryon suppression. The squares are drawn from
a cosmological model and have the same $k$ values and fractional errors
as the Abell/ACO (left) or 2dFGRS (right) surveys. The models
are $\Omega_mh^2 = 0.10$ and $\Omega_b/\Omega_m=0.32$ ({\bf left}) and
$\Omega_mh^2 = 0.11$ and $\Omega_b/\Omega_m=0.18$ ({\bf right}). The
red band is the null hypothesis, where the center corresponds to the
same $\Omega_m$, but $\Omega_b = 0$. This band accounts for uncertainties
in the shape of the power spectrum at small $k$ (see text for details).
The yellow squares are rejected from the null hypothesis for $\alpha = 0.05$.
}

\label{fig::pk_suppress}
\end{figure}

We mimic the $P(k)$ measurements 
using model $P(k)$ at the same $k$-values and fractional errors as given by
Miller and Batuski (2001) and Percival et al. (2001). We also
convolve the model $P(k)$ with the window functions 
of each experiment.
We normalize the nonzero model to the zero baryon model using the power
calculated at three smallest $k$-values of the two different surveys.
This has some consequences that we discuss below.
We allow for uncertainty in this normalization by calculating the
the error-weighted root mean square difference between the
two models over the $k$-values used in the normalization. Ideally, the
zero baryon and non-zero baryon models should have zero difference. But
when an RMS is non-zero, we consider this to be an uncertainty and apply it over
the full range of power in the null hypothesis.
This turns the null hypothesis into
a band of power. We show two examples in Figure \ref{fig::pk_suppress}. On the left,
we show the Abell/ACO $k$-values, errors and window, while on the right we use
the 2dFGRS. Notice that the 2dF $P(k)$ does not probe to large enough
scales (small enough $k$) to get the correct normalization, whereas the Abell/ACO
data does. 

The null hypothesis for our testing procedure is the model $P(k)$ containing
no baryons.
We quantify differences from the null using the False Discovery Rate procedure 
described above.
Instead of looking at the overall shapes of two power spectra, we
approach this problem from a multiple-hypothesis testing point-of-view.
We choose this statistical technique because we want to detect the global
suppression of the $P(k)$ and not simply the features caused by acoustic oscillations.
The FDR technique
allows us to demand that all $P(k)$ measurements above some specified $k_{min}$
are rejected, and not just the bumps and dips.
Since the Abell/ACO and 2dFGRS surveys sample the $P(k)$ at different intervals,
we must account for the sampling differences.
Specifically, we require that at least seven adjacent 2dFGRS data points differ from the
null hypothesis, before we identify a statistical detection of the
baryon suppression. For the more sparsely sampled
Abell/ACO data, we require at least four adjacent data points to differ from the null.
We have chosen these requirements to meet two criteria:
(1) Enough points must be rejected so that we are detecting more than just
the bumps and wiggles
(baryonic acoustic features span very small ranges of $k$--typically one
data point in the Abell/ACO and two in the 2dFGRS); (2) If $\sim 25\%$ of
the total dataset is suppressed, then we have a positive detection.
We note that our results are the same when we make minor changes to 
these requirements.
We use $\alpha = 0.1$ to control false discoveries.

We calculate the probability (for each data point)
that the nonzero power is from the zero baryon power spectrum. We use a two population
test statistic: 
\begin{equation}
z = \frac{\bar{P}_{\rm no ~ baryon} - \bar{P}_{\rm baryon}}{\sqrt{\sigma_{\rm no ~ baryon}^2 + \sigma_{\rm baryon}^2}}
\end{equation}
where $\bar{P}_{\rm no ~ baryon}$ and $\bar{P}_{\rm baryon}$ are the values taken
from the fitting formula of EH98 after inputting $\Omega_m$ and $\Omega_b$ for
a Hubble constant of $70$km s$^{-1}$ Mpc$^{-1}$ and a primordial spectral index,
$n_s = 1$. $\Omega_{\Lambda}$ has no noticable effect on the shape of
the $P(k)$ and the Hubble constant and the spectral index will affect the shapes of
the baryon and non-baryon power spectra in the same way. Therefore, none of these 
affect our test statistic.
All of our models are strictly for Cold Dark Matter universes.
The errors on $\bar{P}_{\rm baryon}$ are the fractional errors
from the original experiments and the error on $\bar{P}_{\rm no ~ baryon}$ is
from the RMS after normalizing the nonzero power spectrum to the zero baryon
power spectrum. We then apply the FDR method as described above.
We use the correlated correction (see Miller et al. 2001c) to account
for the fact that the 2dFGRS $P(k)$ points are correlated. 

To recap, our algorithm is as follows: (1) create model power spectra
with and without baryons over a range of $\Omega_m$ and $\Omega_b$; (2)
convolve these models with the appropriate window function; (3) normalize
the baryonic $P(k)$ to the non-baryonic $P(k)$ for the three smallest
$k$-values in the appropriate survey; (4) apply an error-band to the non-baryonic
$P(k)$ using the RMS difference to the baryonic $P(k)$; (5) use FDR to find
significant differences between the non-baryonic and baryonic
convolved $P(k)$, using the wavenumbers and errors of the appropriate survey.

\begin{figure}
\plottwo{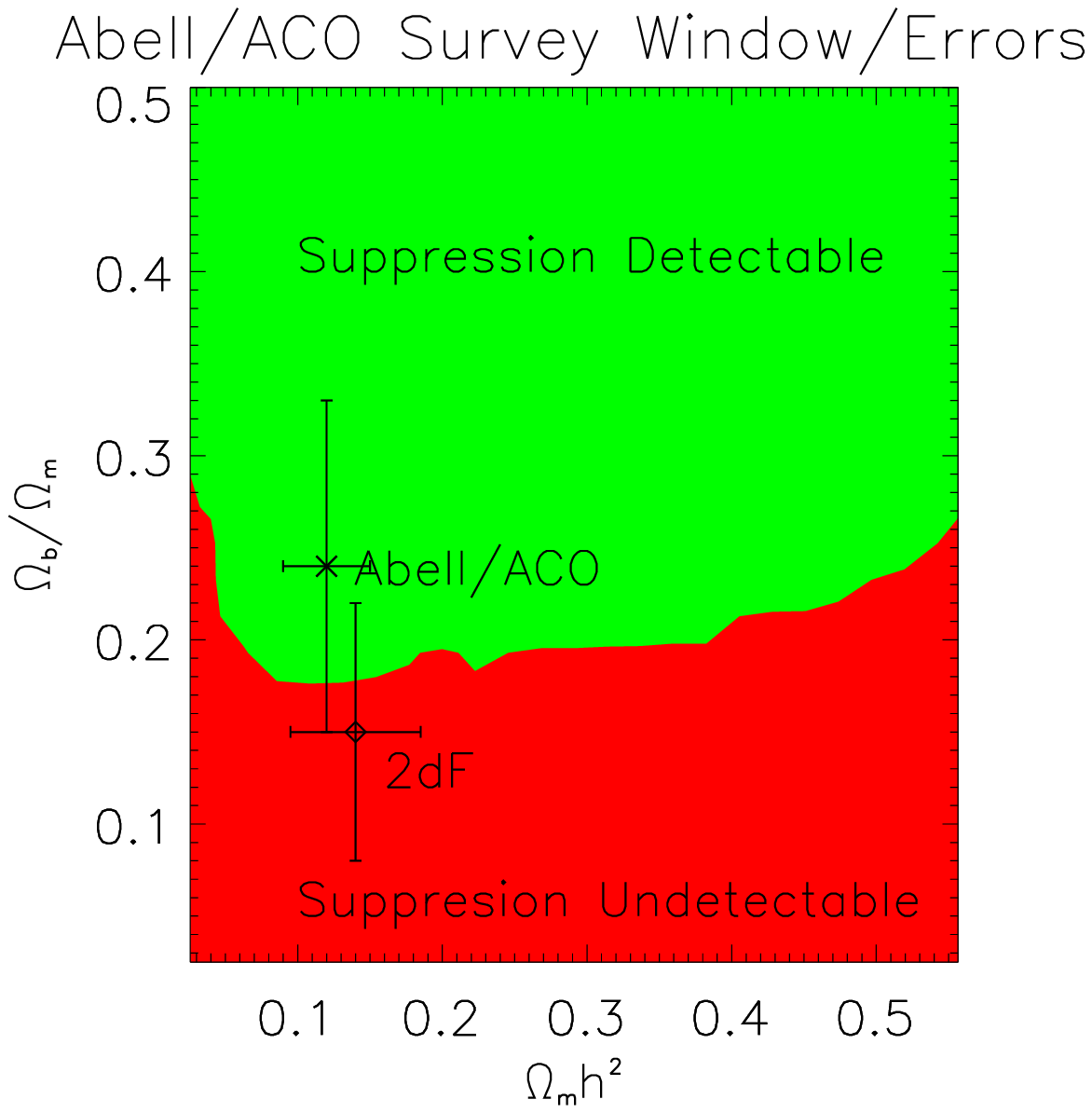}{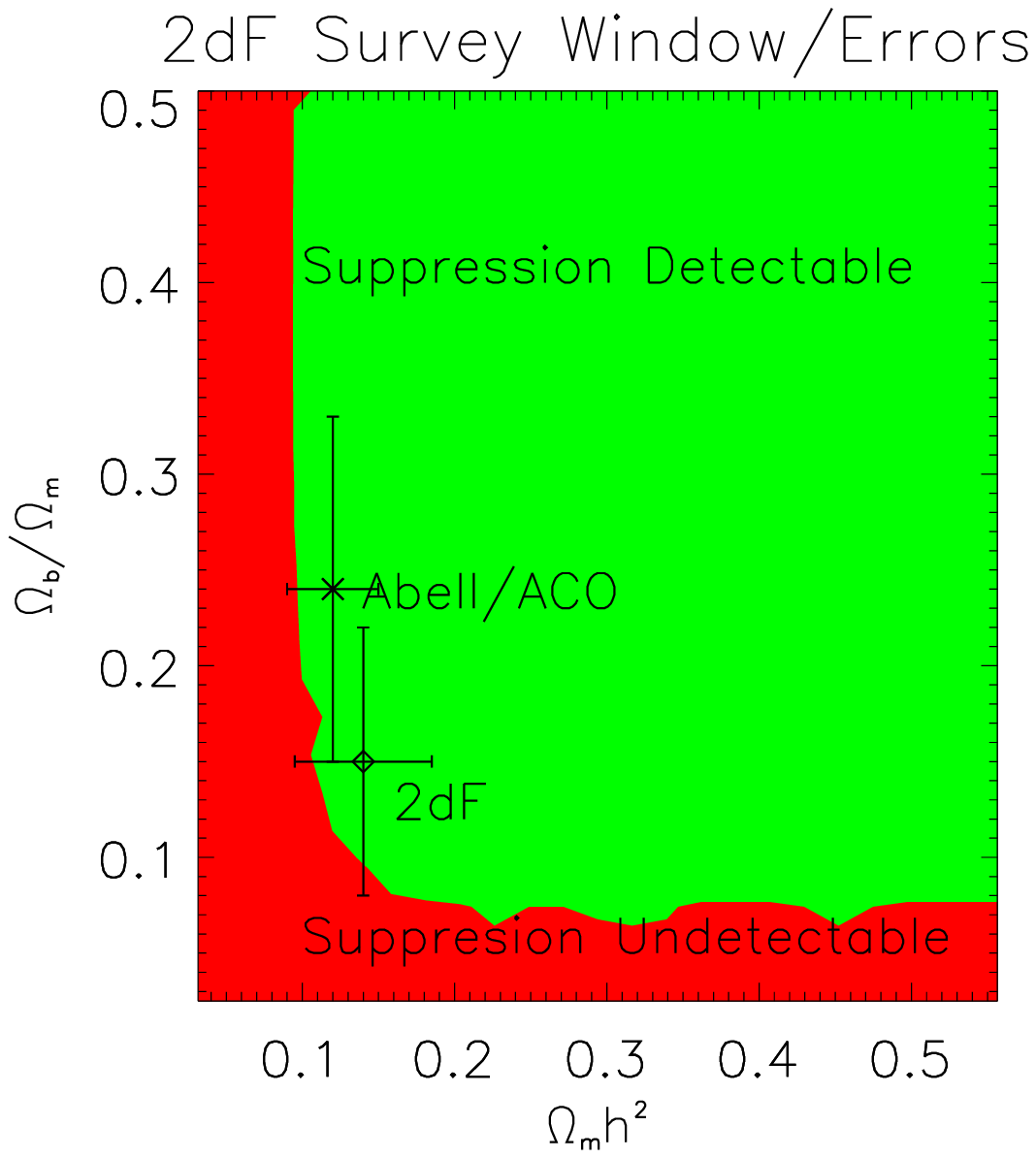}
\caption[]{The regions where the suppression of the power spectrum
due to the baryons can be detected or not. In the left panel, we use
the model power spectra calculated at the Abell/ACO $k$-values, their errors, and convolved
with their window (Miller \& Batuski 2001). On the right, we use those
of the 2dFGRS data (Percival et al. 2001). Notice that the 2dFGRS would not detect 
baryon suppression for small $\Omega_mh^2$ since the power is not calculated
(accurately) to large enough scales. Likewise, the Abell/ACO data cannot
detect baryon suppression for high $\Omega_mh^2$, since they have too few
measured points on small scales. The points are the best fit cosmological
models from Miller, Nichol, and Batuski (2001) and Percival et al. (2001).}
\label{fig::contours}
\end{figure}

In Figure \ref{fig::contours} we present our results. These plots separate
two regions in parameter space: one where the baryon suppression could
have been detected (green), the other where the suppression could not be detected (red). 
The left side edge of non-detectability in the 2dFGRS is directly tied to the availability of
low $k$ measurements. As $\Omega_mh^2$ gets smaller, the sound horizon gets
larger. This means that for small $\Omega_mh^2$, the scale at which
the non-baryonic and baryonic power spectra become identical gets large.
Therefore, one would need measurements of the $P(k)$ on these large scales
to accurately normalize the baryon spectrum to the non-baryon spectrum.
If a survey cannot probe to these large scales,
the normalization occurs over a range in $k$-space where the two power spectra
are not identical, resulting in large uncertainty and the error
band around the null hypothesis gets very wide. 

The lower limits on the  baryon fraction are a direct consequence of the error
bars on the $P(k)$ measurements, i.e. the smaller the error, the lower the
baryon fraction one can detect through suppression. In this sense, the
Abell/ACO and 2dFGRS are complementary. An Abell/ACO type survey
(all-sky but few data) probes to lower $\Omega_mh^2$ while
the 2dFGRS  (small sky coverage but many data) probes to lower baryon fractions.

In both surveys, the reported baryon fractions (by Miller et al. and by Percival
et al.) are within their respective ranges of detectability. The 
lower 2dFGRS baryon fraction (compared to the Abell/ACO) would be
undetectable in the Abell/ACO power spectrum due to its larger errors and
sparse sampling.
We note that both surveys find results for $\Omega_mh^2$ and $\Omega_b/\Omega_m$
that are just in the range of
detectability ($1 \sigma$).
Since the suppression in power from the baryons occurs over a large
range in $k$, it is likely that these surveys are detecting baryon effects through
suppression, rather than through any discrete features.

\section{Suppression via DM Scattering}
Besides baryons, interaction between dark matter and baryons could also 
suppress the power spectrum of the fluctuation. Although for the WIMPs such
effect is too small to be observable (Chen et al. 2001), 
a number of dark matter candidates, especially some associated 
with the strongly self-interacting 
dark matter particles (Spergel \& Steinhardt 2000), 
could have such interaction with baryons (e.g. 
Starkman et al 1990, Wandelt et al. 2000 and references therein). 
If such interaction indeed exists, it would have profound implication for 
the formation and evolution of structure, and its detection could provide 
important clues to the nature of the dark matter. 

The matter power spectrum is suppressed by such interaction, because 
momentum is transfered from the baryon-photon fluid to 
the dark matter in baryon-dark matter collisions (Chen et al 2002). 
The magnitude of this suppression 
depends on both the baryon-dark matter interaction cross section and the 
dark matter mass. We use the same technique as described above to 
look for such suppression, and determine the range of cross section and mass 
for which such suppression could be detected from the Abell/ACO and 2dFGRS data.
In this exercise, we fix $\Omega_mh^2 = 0.2$ and $\Omega_bh^2 = 0.02$.

\begin{figure}
\plottwo{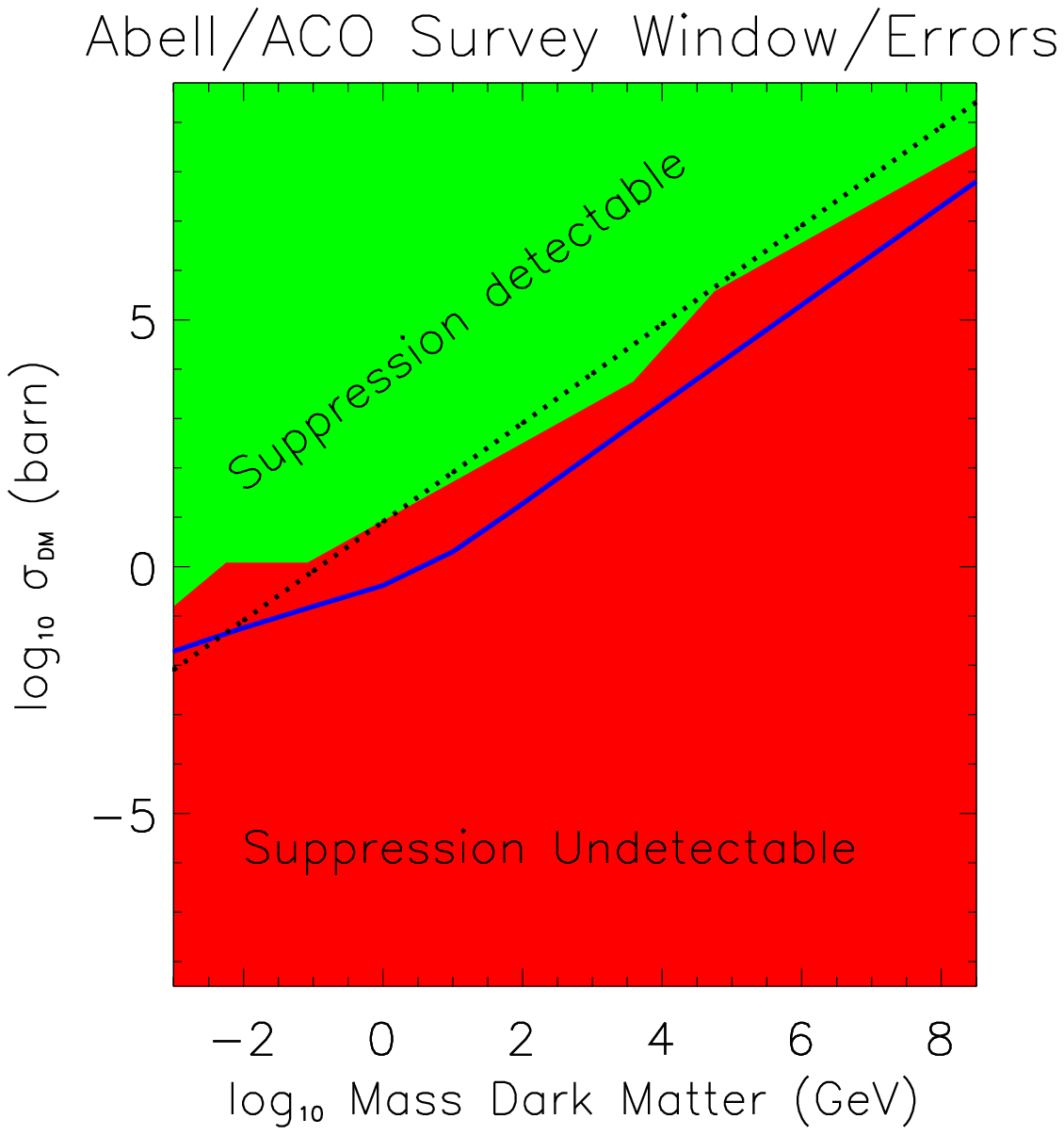}{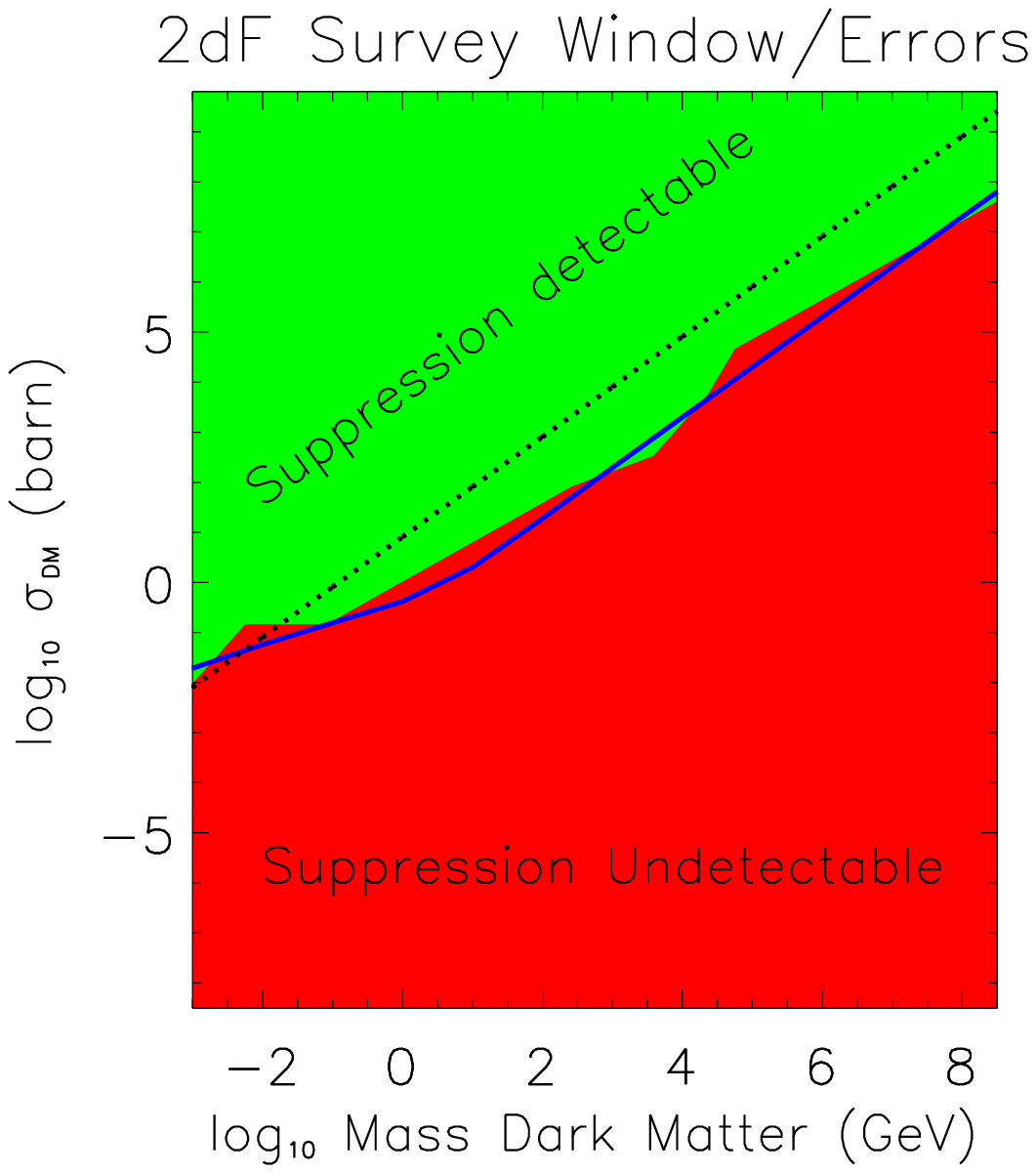}
\caption[]{The regions where the suppression of the power spectrum is
non-negligible due to an interaction between the DM and
the baryons. The x-axis is the mass of a dark mater particle (in GeV) and
the y-axis is the DM-baryon cross-section interaction (in $10^{-24} cm^{-2}$).
In both plots, we compare an $\Omega_mh^2 = 0.2$, $\Omega_bh^2 = 0.02$,
$n_{spectral} = 1$ power spectrum without DM interactions to
the same model with interaction. In the left panel, we use
the power measured at the Abell/ACO $k$-values, their errors, and convolved
with their window (Miller \& Batuski 2001). On the right, we use those
of the 2dFGRS data (Percival et al. 2001). The region above the solid blue line
was ruled out by Chen et al. (2002). The black dotted line is from Spergel and
Steinhardt (2001).}
\label{fig::contours_dm}
\end{figure}

Our result is plotted in Fig. \ref{fig::contours_dm}. The solid blue curve in the plot 
is the limit obtained from combing current CMB and large scale structure 
data by Chen et al 2002. Above this curve, the DM mass and cross-section of the
DM-baryon interaction is ruled out at the 95\% confidence level. 
Using the techniques described in this paper, 
we find that the dark matter-baryon interaction is detectable with
these two survey data sets only if they 
are above or very close to this limit. The detectable range for the 2dFGRS
survey runs parallel to the Chen et al limit, while 
the Abell/ACO data set is seen to be slightly weaker for this purpose. 
This result is not surprising, since the more sensitive 
large scale structure constraint in the Chen et al paper 
is also based on the same 2dFGRS data set. Despite different techniques, 
on this log-log plot the difference appears almost indistinguishable.  
However, Spergel and Steinhardt (2000) suggest a much
stronger DM self-interaction (the dotted line if Figure \ref{fig::contours_dm}).
Thus, if the DM-baryon interaction is 
as strong as DM self-interactions, we should be able to 
detect this suppression with the current 2dFGRS data. 

\section{Summary}

We have examined two large surveys for the effects of baryons on
the luminous matter power spectrum. 
We have shown that the window effects on the measured $P(k)$ are minimal
($<5$\%) for the Abell/ACO survey, while they are significant in the 2dFGRS.
The 2dFGRS window not only smoothes out any features, but it also moves
those features to higher $k$.  We have shown that the features detected
in the Abell/ACO $P(k)$ by Miller, Nichol, and Batuski (2001),
are entirely erased by the 2dFGRS window.
There are features in the 2dFGRS power spectrum visible
to the eye, but they are not significant after a thorough statistical analysis.

After convolving the Abell/ACO power spectrum with the 2dFGRS window, we
find that the shapes of the two power spectra are entirely consistent
from $k = 0.02h$Mpc$^{-1}$ to $k = 0.15h$Mpc$^{-1}$. Neither survey
shows evidence for a turnover in power toward a scale-invariant spectrum.

Since the presence of baryons suppresses the power spectrum on
scales smaller than the sound horizon, we examined whether
either of these surveys could have detected this suppression.
We utilized a multiple hypothesis testing
approach (by controlling the False Discovery Rate) as opposed to
an omnibus test (e.g. a $\chi^2$ test). This allowed us to demand
that the detection of the overall suppression, and not simply
the baryonic features. Since neither the 2dFGRS nor
the Abell/ACO have accurate power measurements on the largest
scales (where there is no baryonic suppression), we 
consider model power spectra which have the same
attributes (windows, sampling, errors, etc) of the two surveys.
We find that the Abell/ACO survey and the 2dFGRS survey are complementary
in that the former can probe to smaller $\Omega_m$ due to its
volume and window, while the latter can probe to smaller $\Omega_b/\Omega_m$
due to its better sampling and smaller errors.

Recently, Chen et al. (2002) had shown that scattering interactions
between dark matter and the baryons will also cause significant
suppression in the matter power spectrum. We fix $\Omega_m$ and
$\Omega_b$ and examine a range of dark matter masses and DM-baryon
cross sections to see whether either of these surveys could detect
this suppression. We find that such interaction is detectable if the 
interaction strength is above or close to the current limit put by
the CMB and large scale structure data. However, we note that if the
interaction strength between the dark matter and baryon is as strong as
the Spergel-Steinhardt 
dark matter self-interaction as speculated in Wandelt et al. (2000),
it should already be detectable.  

\section{Discussion}
Most efforts in this area of research have focused on 
finding the bumps and dips
in the power spectrum resulting from acoustic oscillations (
Eisentstein et al. 1998; Miller et al, 2001a,b;
Meiksin, White, and Peacock 1999). However, even when such
features are found (e.g. Miller et al. 2001a), they may not be the direct result
of the baryons (although statistical tests like FDR can rule out their being caused
by random fluctuations of noise). 
At best, features in the power spectrum can said to be consistent
with those expected from acoustic oscillations.

In this paper, we take a new approach: we study how surveys can detect the
baryons through a global suppression of the power spectrum on scales smaller
than the sound horizon. Such a technique is much more robust to detecting the effects
of the baryons, since it occurs over a large range of scales (as opposed to
the two or three nearly discrete features from acoustic oscillations).
Specifically,  we have examined the
detectability of any suppression in the $P(k)$ on scales $0.015 < k < 0.15$, which 
could be due to the existence of baryons as well as DM-baryon scattering interactions.

The local $P(k)$ contains a wealth of information imprinted on it.
Unfortunately, our measurements of the local $P(k)$ are as yet
unable to fully extract all this information, ({\it e.g.} the
unambiguous detection of baryons). While features in the $P(k)$ have been
found at $k \sim 0.04$ and
$k \sim = 0.1$ in many different (and independent) datasets (Miller et al. 2001a),
there is no conclusive proof that they are the result of baryon fluctuations
(although see Miller et al. 2001b). Likewise, the accuracy of power measurements on
the largest scales is not yet good enough to use the techniques described in
this paper on real $P(k)$ measurements. However, under the hypothesis that
the current data on large-scales can accurately trace the power,
we use model power spectra to find 
that both the Abell/ACO (through its volume) and the 2dFGRS
(through its sampling) are close to being able to detect baryon suppression. 
The combination of volume
and sampling from the Sloan Digital Sky Survey may soon provide the answer
(York et al. 2000, Stoughton et al. 2002).
Therefore, it is important to continue to pursue high precision
measurements of the $P(k)$ over as large a range of $k$'s as possible. This
should be achieved through larger surveys of clusters and galaxies which
possess a well-determined, compact window function and a high sampling
density. Likewise, strong dark matter--baryon interactions could also produce a suppression
in the matter power spectrum, whose evidence could be found using these methods. 
Current data sets could already detect such interaction if  its
strength is as strong as the Spergel-Steinhardt self interaction, and this could
be further improved with the SDSS data. 

{\noindent {\bf Acknowledgments} }
X.C. is supported by the NSF under grant PHY99-07949.

\begin{thebibliography}{}
\bibitem[Balbi et al.(2000)]{2000ApJ...545L...1B} Balbi, A.~et al.\ 2000, 
\apjl, 545, L1 
\bibitem[Benjamini and Hochberg (1995)]{bej} Benjamini, Y., Hochberg, Y., 1995, JRSSB, 57, 289
\bibitem[Chen, Kamionkowski, Zhang (2001)]{ckz01} Chen, X., Kamionkowski, M., 
Zhang, X., 2001, \prd, 64, 021302.
\bibitem[Chen, Hannestad \& Scherrer (2002)]{chs02} Chen, X., Hannestad S., 
Scherrer, R. J., 2002, PRD accepted (astro-ph/0202496).
\bibitem[Colless et al.(2001)]{2001MNRAS.328.1039C} Colless, M.~et al.\ 
2001, \mnras, 328, 1039 
\bibitem[de Bernardis et al.(2002)]{2002ApJ...564..559D} de Bernardis, 
P.~et al.\ 2002, \apj, 564, 559 
\bibitem[(Eisenstein \& Hu 1998)]{eishu98} Eisenstein, D.J., Hu, W. 1998, \apj, 496, 605
\bibitem[Eisenstein, Hu, Silk, \& Szalay(1998)]{1998ApJ...494L...1E} 
Eisenstein, D.~J., Hu, W., Silk, J., \& Szalay, A.~S.\ 1998, \apjl, 494, L1 
\bibitem[Eisenstein \& Hu(1999)]{1999ApJ...511....5E} Eisenstein, D.~J.~\& Hu, W.\ 1999, \apj, 511, 5 
\bibitem[Elgaroy et al. 2002]{elg02} Elgaroy et al. 2002, PRL in press, astro-ph/0204152
\bibitem[Feldman, Kaiser, \& Peacock(1994)]{1994ApJ...426...23F} Feldman, 
H.~A., Kaiser, N., \& Peacock, J.~A.\ 1994, \apj, 426, 23 
\bibitem[Halverson et al.(2002)]{2002ApJ...568...38H} Halverson, N.~W.~et 
al.\ 2002, \apj, 568, 38 
\bibitem[Hamilton, Tegmark, \& Padmanabhan(2000)]{2000MNRAS.317L..23H} 
Hamilton, A.~J.~S., Tegmark, M., \& Padmanabhan, N.\ 2000, \mnras, 317, L23 
\bibitem[Hamilton \& Tegmark (2002)]{2002MNRAS.330L..506H} 
Hamilton, A.~J.~S. and Tegmark, M. 2002, \mnras, 330, 506 
\bibitem[Hopkins et al.(2002)]{2002AJ....123.1086H} Hopkins, A.~M., Miller, 
C.~J., Connolly, A.~J., Genovese, C., Nichol, R.~C., \& Wasserman, L.\ 
2002, \aj, 123, 1086 
\bibitem[Hoyle et al.(2002)]{2002MNRAS.329..336H} Hoyle, F., Outram, P.~J., 
Shanks, T., Croom, S.~M., Boyle, B.~J., Loaring, N.~S., Miller, L., \& Smith, R.~J.\ 2002, \mnras, 329, 336 
\bibitem[Lee et al.(2001)]{2001ApJ...561L...1L} Lee, A.~T.~et al.\ 2001, 
\apjl, 561, L1 
\bibitem[Meiksin, White, \& Peacock(1999)]{1999MNRAS.304..851M} Meiksin, 
A., White, M., \& Peacock, J.~A.\ 1999, \mnras, 304, 851 
\bibitem[Melchiorri et al.(2000)]{2000ApJ...536L..63M} Melchiorri, A.~et 
al.\ 2000, \apjl, 536, L63 
\bibitem[Miller et al.(1999)]{1999ApJ...524L...1M} Miller, A.~D.~et al.\ 
1999, \apjl, 524, L1 
\bibitem[Miller \& Batuski(2001)]{2001ApJ...551..635M} Miller, C.~J.~\& Batuski, D.~J.\ 2001, \apj, 551, 635 
\bibitem[Miller, Nichol, \& Batuski(2001a)]{2001ApJ...555...68M} Miller, 
C.~J., Nichol, R.~C., \& Batuski, D.~J.\ 2001a, \apj, 555, 68 
\bibitem[Miller, Nichol, \& Batuski(2001b)]{2001Sci...292.2302M} Miller, 
C.~J., Nichol, R.~C., \& Batuski, D.~J.\ 2001b, Science, 292, 2302 
\bibitem[Miller et al.(2001c)]{2001AJ....122.3492M} Miller, C.~J.~et al.\ 
2001c, \aj, 122, 3492 
\bibitem[Miller, Nichol, Genovese, \& Wasserman(2002a)]{2002ApJ...565L..67M} 
Miller, C.~J., Nichol, R.~C., Genovese, C., \& Wasserman, L.\ 2002a, \apjl, 
565, L67 
\bibitem[(Miller {\it et al.} 2002)]{mil02b} Miller, C.J., Krughoff, K.S.,  Batuski, D.J., Slinglend, K.A., Hill, J.M., 2002b, AJ in press
\bibitem[(Netterfield et al. 2002)]{net02} Nettefield, C.B., et al. 2002, ApJ in press
\bibitem[Narayanan, Berlind \& Weinberg(2000)]{2000ApJ...528....1N} 
Narayanan, V.\ K., Berlind, A.\ A.\ \& Weinberg, D.\ H.\ 2000, \apj, 528, 1 
\bibitem[Percival {\it et al.} 2001]{percival} Percival, W.\ J. et al,
2001, \mnras, 327, 1297
\bibitem[Saunders et al.(2000)]{2000MNRAS.317...55S} Saunders, W.\ et al.\ 
2000, \mnras, 317, 55 
\bibitem[Spergel \& Steinhardt 2000]{ss00} 
Spergel, D. N., Steinhardt, P. J., 2000, \prl, 84, 3760.
\bibitem[Starkman et al. 1990]{s1990} Starkman, G. D., Gould, A., 
Esmailzadeh, R., Dimopoulous, S., 1990, \prd, 41, 3594.
\bibitem[Stoughton et al.(2002)]{2002AJ....123..485S} Stoughton, C.~et al.\ 
2002, \aj, 123, 485 
\bibitem[Tago et al.(2002)]{2002AJ....123...37T} Tago, E., Saar, E., 
Einasto, J., Einasto, M., M{\" u}ller, V., \& Andernach, H.\ 2002, \aj, 
123, 37 
\bibitem[(Tegmark et al. 2002)]{teg00} Tegmark, M., Hamilton, A.J.S., and Xu, Y. 2002, MNRAS in press
 astro-ph/0111575
\bibitem[Wandelt et al. 2000]{wan00} Wandelt, B.D. et al., 2000, 
{\it Proceedings of Dark Matter 2000}, [astro-ph/0006344].
\bibitem[Wang et al. 2001]{wang01} Wang, X., Tegmark, M. and Zaldarriaga, M. 2001,
  Phys. Rev. D. submitted, astro-ph/0105091 -- WTZ
\bibitem[York et al.(2000)]{2000AJ....120.1579Y} York, D.\ G.\ et al.\ 
2000, \aj, 120, 1579 
\end{thebibliography}
\end{document}